# Inversion of Sea Ice Spectral Albedo to Estimate Under-Ice Transmittance

C. Perron[1,2], B. Raulier[1], P. Massicotte[1], M. Vancoppenolle[2] and M. Babin[1]

[1]CNRS – Université Laval – Sorbonne Université – International Research Laboratory Takuvik

Département de biologie et Québec-Océan, Université Laval, Québec, QC, Canada.

[2]Sorbonne Université, LOCEAN-IPSL, CNRS/IRD/MNHN, Paris, France.

Corresponding author: Christophe Perron (chper110@ulaval.ca)

**Key Points:**

- From the ground at local scale, spectral albedo inversion can estimate vertically resolved scattering properties of snow-covered sea ice.

- Vertically resolved scattering properties can be used to estimate transmittance with improved accuracy in comparison to current methods.

- If applied to remote sensing, spectral albedo inversion could improve under-ice photosynthetically active radiation estimates at pan-Arctic scale

**Abstract**

Sunlight radiation under snow-covered sea ice obtained from remote sensing could help assess under-ice primary production at pan-Arctic scale. Yet, the current remote sensing methods to estimate sunlight transmittance under sea ice is limited by its reliance on imprecise snow depth products and its inability to sense microstructure-driven variations in snow and ice light scattering properties. Based on Monte-Carlo simulations of radiative transfer, we developed an inversion method relying solely on spectral albedo to estimate transmittance under snow-covered sea ice. The method analyses albedo spectral information to derive the vertically resolved scattering properties of snow and sea ice above the freeboard. Assuming fixed columnar ice physical and optical properties, transmittance is then estimated. At ground level, our spectral albedo inversion method is more precise than the current approaches. We argue this is because it implicitly accounts for the variability in snow scattering properties. This method could significantly improve the satellite estimation of photosynthetically available radiation under sea ice, especially because it does not need snow depth.

**Plain Language Summary**

**This study presents a new method with the potential to estimate using satellites how much light passes through sea ice. Understanding under-ice light is important because it helps predict when and where during the spring season marine life starts to grow in the Arctic ocean. The existing methods cannot provide under-ice light with enough precision because it needs to assess snow depth from space, which is very challenging. Also, it does not consider how the ever-evolving microstructure of snow and sea ice might influence light propagation. This alternative method analyses how the different colors constituting sunlight are reflected when they hit the snow surface. What makes this method unique is its ability to extract precise inherent snow and sea ice structural information without needing snow depth data. When tested from the ground, this novel approach proves more precise than the existing one. With no need for independent snow depth measurement, this approach could significantly improve satellite estimation of light availability under sea ice.**

1 Introduction

Photosynthetically active radiation (PAR) is a key factor in assessing primary production under Arctic sea ice (Horner & Schrader, 1982). In particular, the drastic increase of PAR in spring dictates the timing of sea ice and under-ice algal blooms (Castellani et al. 2017). This PAR threshold, which triggers phytoplankton bloom, is occurring increasingly early in the season (Ardyna et al., 2014; Arrigo & van Dijken, 2011). Given these changes, precise and frequent remote sensing maps of under-ice PAR could help assess when and where under-ice primary production will begin on a pan-Arctic scale.

Essentially, remote under-ice PAR requires estimating visible sunlight transmittance through snow and sea ice. To estimate transmittance, the current method tested by Stroeve et al. (2021) relies on the two-layer Beer-Lambert approach (Maykut and Untersteiner 1971, Grenfell and Maykut 1977). In this approach, transmittance is obtained by subtracting (1) the fraction of light backscattered to space using broadband albedo and (2) subtracting the fraction of light attenuated inside, using mainly snow and sea ice geometric thicknesses.

This approach faces two difficulties. First, obtaining snow geometric depth from space is challenging. As mentioned by Stroeve et al. (2021) , the uncertainties on remote snow depth, obtained from a combination of modeling with atmospheric reanalyses and satellite drift datasets (Liston et al., 2020), prevents precise quantitative under-ice light assessments. In particular, uncertainties are too high to assess when and where PAR reaches sufficient levels to trigger phytoplankton blooms. Second, the variability of snow and sea ice scattering properties, induced by variability of the microstructure (Kokhanovsky and Zege 2004, Light et al. 2004) are not directly considered. They are indirectly accounted for in broadband albedo (for reflective losses) and classification of snow as dry or wet (for attenuation losses). These scattering properties, together with geometric thickness, are inherently driving radiative transfer in sea ice. Yet, the extent to which fully accounting for their variability could improve transmittance estimates is unknown.

Instead of relying on remotely sensed geometric snow depth to estimate attenuation, the vertically resolved scattering properties could potentially be used. They could be retrieved from information carried in the visible albedo spectrum. Albedo signals reach varying depths in snow and sea ice depending on the wavelength. Thus, the wavelength-by-wavelength analysis of its spectrum can resolve a unique vertical profile of the scattering properties of snow and sea ice above the freeboard. This information is sufficient, in theory, to estimate transmittance under first-year snow-covered sea ice in most scenarios.

Spectral albedo measured, in combination with other structural and optical measurements, has already been used several times to assess the scattering properties of snow and sea ice(e.g. Ehn et al. 2008, Light et al. 2008, Xu et al. 2012, Light et al. 2015), but, it has never been used alone in an automated approach and has never been used to estimate transmittance. Such an automated technique relying on spectral albedo is currently used as a non-destructive diagnostic tool in the medical field. The method estimates the vertically resolved inherent optical properties of multi-layered human tissue to assess their state (Yudovsky and Pilon 2011, Sharma et al. 2014, Hennessy 2015).

In this work, we test whether an automated spectral albedo inversion could estimate the vertically resolved scattering properties of snow and sea ice above the freeboard. Then, we assess whether these vertically resolved scattering properties improve transmittance estimates in comparison to the current Beer-Lambert method, which relies mainly on snow geometric thickness. The validation of the algorithm outputs—vertically resolved scattering properties and transmittance— and the comparison to the Beer-Lambert method are done using measurements obtained over five field campaigns on snow-covered first-year sea ice.

## 2 Methods

We developed an algorithm that compares spectral albedo measurements to Monte Carlo simulations to estimate the vertically resolved scattering properties of snow-covered first-year sea ice and its transmittance . The Monte Carlo simulations which are based on a three-layer radiative transfer model are described in Section 2.1. The algorithm is described in Section 2.2. The field data used for the validation of the algorithm outputs and its comparison to the Beer-Lambert (BL) method is described in Section 2.3.

### 2.1 Monte Carlo Simulations Using a Three-Layer Radiative Transfer Model

We employ Monte Carlo simulations with a three-layer radiative transfer model to represent the relationship between sea ice microstructure and light propagation in sea ice. In the Monte Carlo method, the trajectories of numerous photons through a given medium is simulated numerically based on laws defined by the radiative transfer theory (Chandrasekhar, 1960). In each simulation $N=10^4$ photons were simulated and launched at a zenithal angle $\theta_s$ of 55°. The ratio of launched photons to photons received at a specific location in the medium are used to calculate simulated apparent optical properties such as albedo and transmittance.

Each layer in our numerical environment is defined by three properties: the absorption coefficients $a$ (m-1), the reduced scattering coefficients $b'$ (m-1) and the geometric thickness $h$. The value of $a$ is the only one which depends on wavelength. Thus, $a$ is used to tune the wavelength (see Section 2.1.1). The value of b' is dependent on the state of the microstructure. The value of b' and h for each layer is thus varied to represent the complete variability induced by the vertical microstructure profile of first-year sea ice (see Section 2.1.2).

#### 2.1.1 Absorption Coefficient to Define the Wavelength

The absorption coefficient $a$ (m$^{-1}$) defines the spatial rate at which photons are attenuated along their path in a given medium, a process which is dependent on the photons wavelength $\lambda$ (nm). Thus, by shifting $a$ over 52 increments, we could simulate light spectra in the visible range.

The relation between $a$ and $\lambda$ is based on the following equation:

$$a = \beta \cdot \left(vf \cdot a_{ice}(\lambda) + c_{imp} \cdot a^*_{imp}(\lambda)\right), \qquad (1)$$

Where the absorption amplification factor $\beta$ was set to 1.37 as a compromise between dry and wet snow (Robledano et al., 2023), $vf$ (-) is the ice volume fraction of the given layer, $c_{imp}$ is the absorbing impurities concentration set to 0.72 (g/m³) and $a^*_{imp}(\lambda)$ (m²/g) is the specific absorption coefficient of a mix of mineral dust and soot representing a coastal area (Verin, 2019). The absorption spectrum of pure ice $a_{ice}(\lambda)$ was based on observations from Warren and Brandt (2008). Given the $c_{imp}$ we chose, the spectral range covered $\lambda$ from 435 nm to 940 nm with 37 increments. It is to be noted that this range and the number of increments in it would change for a different $c_{imp}$. For $c_{imp}$ =0 g/m³, the spectral range would cover $\lambda$ from 480 nm to 940 nm with 52 increments.

The value of $a$ had a different value for each layer because of different air volume fraction, but all layers' absorption coefficients varied coherently. The value of $a$ for Layer 1 varied over 52 increments from 0.01 m⁻¹ to 5 m⁻¹. The step between increments was not even and was meant to maximize the spectral resolution at wavelengths where albedo fluctuates most significantly.

### 2.1.2 Reduced Scattering Coefficient to Represent the Microstructure of the Layers

Our Monte Carlo numerical environment idealized the snow-sea ice system as three homogeneous layers (L1, L2 and L3), each with defined reduced scattering properties and geometric thickness $h$. L1 and L2 properties were varied to build a lookup table with properties covering the range of first-year sea ice (see Table 1).

Table 1: Layers parameters of 3D Monte Carlo simulations used to obtain lookup table . Parameters are air volume fraction $vf$, reduced scattering coefficient $b'$, phase function asymmetry parameter $g$, refractive index $n$ and geometric thickness $h$.

| Layer | Level of scattering | $vf$ (-) | $a$ (m⁻¹) | $b'$ (m⁻¹) | $g$ (-) | $n$ (-) | $h$ (m) |
|---|---|---|---|---|---|---|---|
| 1 | high | 0.420 | 0.01-5 | 20-500 | 0.85 | 1 | 0.01-0.2 |
| 2 | medium | 0.31 | 0.01-5 | 5-150 | 0.85 | 1 | 0.1 |
| 3 | low | 0.920 | 0.01-5 | 2 | 0.98 | 1 | 1 |
| 4 | water | - | 0.01 | 0.001 | 0.9 | 1 | 1 |

The scattering properties can be approximated by a single parameter named the reduced scattering coefficient $b'$ (m⁻¹). The value of $b'$ is equal to $b(1-g)$, where the scattering coefficient $b$ (m⁻¹) describes the spatial rate at which photons are deviated, and the asymmetry parameter g (-) is the mean cosine of the Henyey-Greenstein phase function describing photons angular redirection.

The range of $b'$ and h of each layer were informed from the physical structure meant to be represented:

**Layer 1** (L1) is meant to represent high scattering snow. Both its depth and scattering properties vary.

**Layer 2** (L2) is meant to represent medium scattering transformed snow and/or sea ice above the freeboard. Its scattering properties vary. The value of $h_2$ was fixed to 0.1 m in the case of this study. Yet, the algorithm offers to the user the ability to set $h_2$ between 0.05 m and 0.3 m.

**Layer 3** (L3) is meant to represent low scattering columnar ice below the freeboard. Its depth and scattering properties were fixed to $h_3$ =1 m and $b'_3$=2 m$^{-1}$ to limit the size of the lookup table. For first-year ice, the role of L3 in attenuating light is not as important as for L1 and L2 because $\tau'_{1,2}$ is much larger than $\tau'_3$ in most scenarios.

## 2.2 Spectral albedo inversion algorithm

The $\alpha_\lambda$ -inversion algorithm takes measured ground spectral albedo $\alpha_{\lambda,mes}$ (-) as input. It compares this spectrum to every albedo simulated spectrum $\alpha_{\lambda,sim}$(-) contained in a lookup table built from Monte Carlo simulations described in Section 2.1. Every $\alpha_{\lambda,sim}$(-) is associated to a unique vertically resolved $b'$ profile (defined by $b'_1, h_1, b'_2$ ) in the lookup table. The simulation that best fits $\alpha_{\lambda,mes}$ defines vertically resolved $b'$ profile and the corresponding simulated under-ice spectral transmittance $T_{\lambda,inv}$ (-) (see Figure 1). The complete process is described below:

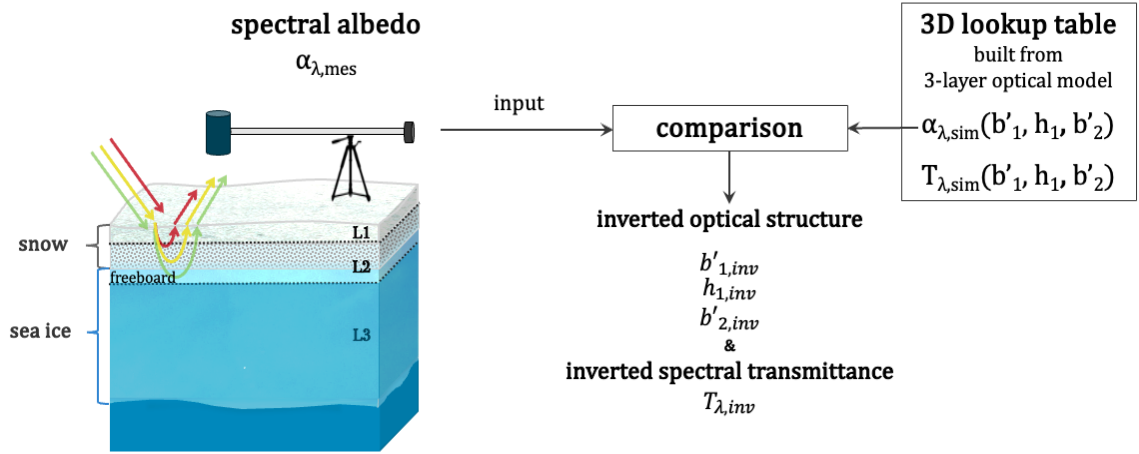

Figure 1 : The automated $\alpha_\lambda$ -inversion algorithm uses spectral albedo ground measurement to invert snow-sea ice vertical optical structure and spectral transmittance under-sea ice. In our three-layer model, layer 1 and 2 properties are assumed to vary. Layer 3 properties are assumed constant.

**Input**

The only input of the algorithm is:

- Measured spectral albedo $\alpha_{\lambda,mes}$: $\lambda$ from 532-700 nm.

It is to be noted that the $\lambda$ range chosen here does not correspond to the spectral range of the Monte Carlo simulations. This shorter range was chosen to optimize the inversion process and can be changed by the user as long as $\lambda$ values are within the spectral range of the Monte Carlo simulations.

**3D Lookup Table**

The lookup table is a 3D parameter space. Each element of this 3D space contains two spectra simulated by Monte Carlo using a three-layer sea ice optical model:

- Simulated spectral albedo $\alpha_{\lambda,sim}$: $\lambda$ from 435 – 940 nm, 37 points.

- Simulated spectral transmittance $T_{\lambda,sim}$: $\lambda$ from 435 – 940 nm, 37 points.

The 3 dimensions are:

- Layer 1 reduced scattering coefficient $b'_1$ : 14 points from 20-500 m$^{-1}$.

- Layer 1 thickness $h_1$: 6 points from 0.01-0.2 m.

- Layer 2 reduced scattering coefficient $b'_2$ : 8 points from 5-150 m$^{-1}$.

The resolution of all three parameters is multiplied by 10 using linear interpolation within the algorithm. Layer 2 depth, $h_2$, is assumed to be 0.1 m and layer 3 properties were kept constant to limit the size of the table.

**Comparison**

The input $\alpha_{\lambda,mes}$ is compared to each simulated spectrum $\alpha_{\lambda,sim}(b'_1, h_1, b'_2)$ in the 3D parameter lookup table by minimizing error $\chi$ :

$$\chi(b'_1, h_1, b'_2) = \sum_{\lambda=532\,nm}^{700\,nm} \frac{\left(\alpha_{\lambda,mes} - \alpha_{\lambda,sim}(b'_1, h_1, b'_2)\right)^2}{\alpha_{\lambda,mes}}. \qquad (2)$$

The smallest error $\chi$ provides the best matching layer properties $b'_1, h_1, b'_2$, then used as coordinates pointing to the best-matching simulation in the 3D lookup table.

The best matching simulation provides diagnostic simulated transmittance spectrum $T_{\lambda,sim}$ also contained in the 3D parameter space.

**Outputs**

- Vertical optical structure: $b'_{1,inv}$ (m⁻¹), $h_{1,inv}$ (m), $b'_{2,inv}$ (m⁻¹), wavelength-independent.

- Transmittance spectrum $T_{\lambda,inv}$ (-): $\lambda$ from 435 - 940 nm, 37 points.

Note that the spectral range of simulated AOPS is specific to the level of absorbing impurities we chose for this experiment. The spectral range is determined from absorption coefficient range and 
$a=\beta \cdot \left(vf \cdot a_{ice}(\lambda) + c_{imp} \cdot a^*_{imp}(\lambda)\right)$, (1. The spectral range and the number of increments vary with the level of impurities.

For analysis, the inverted vertical optical structure of layer 1 and 2 were combined into a single parameter called the reduced optical depth $\tau'_{1,2}(-)$:

$$\tau'_{1,2} = b'_{1,inv} \cdot h_{1,inv} + b'_{2,inv} \cdot 0.1\ m,\ (3)$$

Which is well correlated to under-ice transmittance.

Inverted broadband transmittance $T_{inv}$ was obtained by integrating $T_{\lambda,inv}$ over $\lambda$ between 485 nm and the highest wavelength available from both the simulation (940 nm) and the measurement (sensor-dependent):

$$T_{inv} = \frac{\sum_{i=1}^{n} T_{\lambda,inv\ i} \cdot E_{\lambda,0+\downarrow} \cdot \Delta\lambda_i}{\sum_{i=1}^{n} E_{\lambda,0+\downarrow} \cdot \Delta\lambda_i},\quad (4)$$

Where $E_{\lambda,0+\downarrow}$ (W/m²) is the downwelling spectral irradiance at the surface.

**2.3 Field Data**

Our algorithm was validated using measurements from five campaigns on Arctic first-year snow-covered sea ice. These five campaigns covered different regions of the Arctic and time periods, and the measurements were made by different operators each with their own standards. Various multispectral or hyperspectral sensors were used (see Table 2).

Table 2 : List of missions and sensors used to validate vertically resolved reduced scattering coefficient b' and under-ice transmittance T.

| Mission | | | | | | | | Sensor | | To Validate: | |
|---|---|---|---|---|---|---|---|---|---|---|---|
| Name | Year | Location | Season | # Obs | Snow Depth (m) | Freeboard (m) | Ice Thickness (m) | Type | λ Range (nm) | b' | T |
| QikIce | 2023 | Baffin Bay | spring | 31 | 0.18±0.09 | 0.08±0.05 | 1.2±0.2 | C-OPS | 380:875 19 bands | x | x |
| Baysis | 2018 | Hudson Bay | summer | 7 | 0.03±0.05 | 0.06±0.07 | 1.1±0.6 | Trios | 400:1:700 | | x |
| Green Edge | 2015 2016 | Baffin Bay | spring spring | 24 17 | 0.3±0.1 0.37±0.09 | - - | 1.21±0.09 1.39±0.09 | MAYA 2000 | 200:0.5:1120 | | x |
| SHEBA | 1998 | Beaufort sea | spring& summer | 6 | 0.006±0.01 | - | 1.2±0.6 | Analytical Spectral Devices | 400:0.5:800 | | x |

**2.3.1 Measured Spectral Albedo**

The core measurement was spectral albedo $\alpha_{\lambda,mes}$ (-) in the visible range. For each campaign, a sensor with $n$ wavebands $\lambda_{1..n}$ was positioned on a horizontal pole mounted on a tripod. The ratio of downward $E_{\lambda,0+\downarrow}$ (W/m²) to upward $E_{\lambda,0+\uparrow}$ (W/m²) spectral irradiance was used to obtain:

$$\alpha_{\lambda_{1..n},mes} = \frac{E_{\lambda_{1..n},0+\uparrow}}{E_{\lambda_{1..n},0+\downarrow}}. \qquad (5)$$

**2.3.2 In Situ Vertically Resolved Scattering Properties**

To validate snow and sea ice vertically resolved scattering properties retrieved from albedo inversion, we measured *in situ* the vertical profile of the reduced scattering coefficient $b'_{in\ situ}$ ( m⁻¹) at 16 sites in Qikiqtarjuaq, Nunavut, Canada at the beginning of May during the QikIce2023 campaign. These vertical profiles were obtained with an active diffuse reflectance probe developed and validated in the field for the study of sea ice (Perron et al., 2021).

The measurement principle is the following: the probe injects red laser light at $\lambda$=633 nm within the medium and measures the backscattering at various lateral distances up to 20 mm away from the laser using receiving optical fibres. Comparison to Monte Carlo simulations allows to estimate $b'_{in\ situ}$ for a localized snow or sea ice volume of roughly 1 dm³. At the snow topmost surface, $b'_{in\ situ}$ was obtained with the probe leaned vertically on the snow surface. Efforts were made not to compress the snow. Then, a two-inch hole was drilled with an auger (Kovacs Entreprises™, Roseburg, United States), keeping the snow cover as intact as possible. Punctual measurements were obtained with the probe looking sideward starting 10 cm under the snow surface and then every 10 cm until the bottom of the ice. $b'_{in\ situ}$ were not considered if measured and inferred simulated spatially resolved diffuse reflectance varied by more than 60% at 3 or more detecting fibres positions.

$b'_{1,inv}$ values were compared to in situ $b'_{in\ situ}$ measured at the surface. Because active probe used to measure $b'_{in\ situ}$ senses approximately 0.1 m deep into sea ice (Perron et al., 2021), $b'_{1,inv}$ was adjusted to be coherent to in situ measurements as follows:

$$if\ h_1 < 0.1\ m,\ \ b'^{*}_{1,inv} = \frac{b'_{1,inv} \cdot h_1 + b'_{2,inv} \cdot (0.1 - h_{1,inv})}{0.1}, \quad (6)$$

$$else\ if\ h_1 \geq 0.1\ m, \quad b'^{*}_{1,inv} = b'_{1,inv}. \quad (7)$$

$b'_{2,inv}$ and $b'_{3,inv}$, the latter fixed to 2 m$^{-1}$, were compared to sideward $b'_{in\ situ}$ measurements at depths coinciding with their respective inverted layer. It is to be noted that while $b'_{2,inv}$ and $b'_{3,inv}$, are layer averages, $b'_{in\ situ}$ is a localized measurement for a volume of approximately 1dm$^3$. Thus, both measurements are not exactly representative.

### 2.3.3 Measured Transmittance

For validation, measurements of spectral transmittance, $T_{\lambda,mes}$ (-), were made by inserting an L-arm inside an auger-drilled hole. The ratio of under-ice $E_{\lambda,bot-\downarrow}$ (W/m$^2$) to surface $E_{\lambda,0+\downarrow}$ (W/m$^2$) was obtained for every waveband $\lambda_i$ following:

$$T_{\lambda_i,mes} = \frac{E_{\lambda_i,bot-\downarrow}}{E_{\lambda_i,0+\downarrow}}, \quad (8)$$

and used to calculate the broadband transmittance $T_{mes}$ (-). $T_{mes}$ (-) was integrated between 480 nm and the highest wavelength available both from simulation (940 nm) and measurement (sensor dependent):

$$T_{mes} = \frac{\sum_{i=1}^{n} E_{\lambda_i,bot-\downarrow} \cdot \Delta\lambda_i}{\sum_{i=1}^{n} E_{\lambda_i,0+\downarrow} \Delta\lambda_i}. \quad (9)$$

The choice to start the integration at 480 nm is technical. The spectral range in our simulations depends on the chosen absorbing impurities concentration. While this concentration is fix in this study, the algorithm gives the choice of the concentration to the user. Because we want transmittance integration to start at the same wavelength no matter the choice of impurity, we chose to always start the integration at 480 nm

**2.3.4 Transmittance Modeled from Beer-Lambert Approach**

For comparison, modeled broadband transmittance $T_{mod}$ (-) was calculated from measured snow depth $h_s$ (m) and broadband albedo $\alpha_{mes}$. A Beer-Lambert (BL) attenuation formula similar to Maykut and Untersteiner (1971) was used:

$$T_{mod} = (1 - \alpha_{mes}) \cdot e^{-k_s \cdot h_s - k_i \cdot h_i}. \qquad (10)$$

Attenuation coefficient $k$ was set to 7 or 10 m$^{-1}$ (wet/dry) for snow and to 1 m$^{-1}$ for sea ice, as in Stroeve et al. (2021). The $i_o$ parameter representing the transmittance through a highly attenuating surface layer was not considered because it significantly decreased the performance of the model.

Sea ice thickness $h_i$ (m) was fixed to 1 m to have a fair comparison with the inversion algorithm, in which properties of sea ice under freeboard (L3) are fixed. The aim of this specific analysis is to see how the inverted optical structure of snow and sea ice above the freeboard (L1-L2) could replace snow height remote estimates. Thus, considering actual sea ice thickness would only add complexity to this comparison.

A validation in which transmittance was corrected to meet actual sea ice attenuation is still presented in Appendix B. In this case, a BL correction factor $e^{-k_i \cdot \Delta h_i}$ is used for both estimation methods.

**3 Results**

In this study, we aim to show that spectral albedo contains information on vertically resolved scattering properties of snow and sea ice above the freeboard which can be used to improve transmittance estimate in comparison to current methods. To demonstrate that claim, In situ spectral albedo measurements $\alpha_{\lambda,sim}$ were inverted using a novel algorithm called $\alpha_\lambda$-inversion. The outputs, vertically resolved scattering properties and transmittance, are validated to in situ measurements. The performance of the output transmittance is also compared to the state-of-the-art BL method, for which attenuation is determined from measured snow depth. The importance of microstructure-driven scattering properties on determining transmittance is then assessed.

**3.1 Snow-Sea Ice Vertically Resolved Scattering Properties from $\alpha_\lambda$-inversion**

Our first step is to assess whether vertically resolved scattering properties from $\alpha_\lambda$-inversion algorithm is consistent with in situ scattering properties measured with a localized active probe.

Figure 2 illustrates the validation of layer-resolved $b'_{inv}$ obtained from $\alpha_\lambda$-inversion during the QikIce 2023 campaign. For each site, the inverted reduced scattering coefficient $b'_{inv}$ of each layer is compared to depth-resolved $b'_{in\ situ}$ measurements falling within the depth range of the specific layer. All $b'_{in\ situ}$ measurements falling into a given layer are shown. No layer averaging were performed for $b'_{in\ situ}$ given the discontinuity of the vertical profiles. The axis range spans the documented range of $b'$ for sea ice and snow (Ehn et al. 2008, Light et al. 2008, Light et al. 2015, Perron et al. 2021, Vérin et al. 2022).

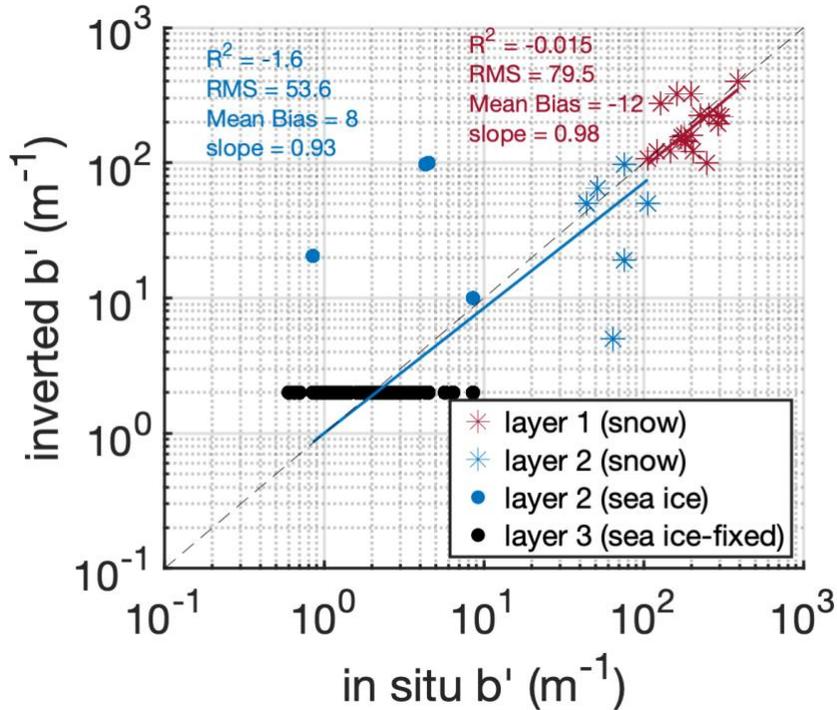

Figure 2: Field validation of vertically resolved scattering properties from spectral albedo inversion (QikIce 2023). The reduced scattering coefficient $b'_{inv}$ estimated for each layer is compared to depth-resolved $b'_{in\ situ}$. The $b'_{inv}$ of layer 3 is fixed to 2 m⁻¹ in our model.

As seen on Figure 2, a distinction in the magnitude of scattering can be observed between layers 1 and 2. This decrease is consistent with in situ measurements. For layer 1, $b'_{1,in\ situ}$ is between 100-400 m⁻¹. For layer 2, $b'_{2,in\ situ}$ is between 40-100 m⁻¹ when probing snow and between 0.7- 9 m⁻¹ when probing sea ice. It suggests that the segmentation of the layers, decided by the algorithm, is driven by an actual decrease in scattering.

This decrease in scattering between layer 1 and 2 is also consistent with field observations. Indeed, at every site, we observed a topmost layer consisting of dry white snow—a mix of precipitated particles, rounded grains, and/or faceted crystals. In most cases, underneath that layer, we found a less white layer of depth hoar with coarser, transformed grains. Coarser grains typically have lower scattering properties (Kokhanovsky & Zege, 2004). In some cases, when the snow cover was thin, the topmost layer lay directly on the sea ice surface.

The lower agreement between $b'_{inv}$ and $b'_{in\,situ}$ for layer 2 can be explained by the mixed composition of layer 2. Layer 2 consists of either transformed snow, a mix of transformed snow and sea ice, or sea ice only. This can be better visualized from $b'$ vertical profiles of all 16 sites provided in Appendix A. These profiles show that the bottom of layer 2 closely coincides with either snow-sea ice interface or the freeboard, as determined by the algorithm. Yet, because layer 2 can consist of snow and sea ice simultaneously, the comparison of in situ and inverted $b'_2$ is potentially problematic. At the sites where layer 2 was a mix of snow and sea ice, $b'_{2,inv}$ represents an average of both media while $b'_{2,in\,situ}$ is either measured in snow or sea ice, but not in a mix of both. Unfortunately, the vertical resolution (10 cm) and the discontinuity of vertical profiles means we can't obtain a layer average for the in situ measurement.

### 3.2 Evaluation of Transmittance from $\alpha_\lambda$-inversion

Our second step is to evaluate transmittance from $\alpha_\lambda$-inversion by comparison to in situ measurements and to the BL method. To this end, Figure 3 compares estimations of transmittance (a) from $\alpha_\lambda$-inversion and (b) from the BL method. They are both evaluated against measured transmittance obtained in the field with radiometers. We obtained estimations from five campaigns on first-year snow-covered Arctic sea ice.

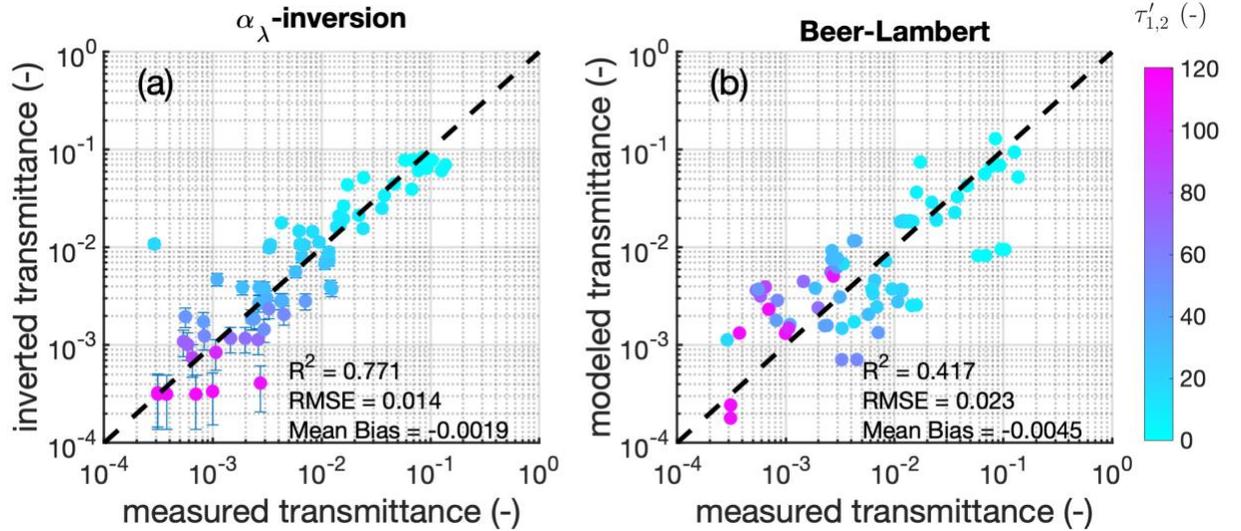

Figure 3: Field validation of transmittance under snow-covered sea ice estimated from (a) inversion of spectral albedo and (b) two-layer Beer-Lambert method using broadband albedo and snow depth. For (a), Layer 3 has a fixed value, while for (b) sea ice thickness has a fixed value. The colorbar is the reduced optical depth of layers 1 and 2 obtained from inversion of spectral albedo. The errorbars represents the stochastic uncertainties on Monte Carlo simulations.

The comparison between both panels of Figure 3 suggests that the $\alpha_\lambda$-inversion method agrees better with observational retrievals than the Beer-Lambert calculation . Indeed, we note a higher R-squared ($R^2$), lower root mean square error (RMSE) and lower mean bias with the spectral-albedo inversion. The BL method displays a greater number of points that are significantly off the linear trend, especially at low and high transmittances.

The better performance of $\alpha_\lambda$-inversion could be explained by its ability to account for some of the variability in the vertically resolved properties of layers 1 and 2. As seen on Figure 3 a, transmittance decays consistently with the augmentation $\tau'_{1,2}$. The value of $\tau'_{1,2}$, combines the effect of both geometric depth and scattering properties and is calculated from vertically resolved scattering properties. On the other hand, the BL method, as displayed here, relies mainly on snow geometric depth to estimate attenuation and, thus, transmittance in sea ice. The attenuating effect of snow scattering properties, hidden in the attenuation coefficient $k_s$, depends only on the binary state of the snow (wet or dry). The lack of sensitivity of the BL method to variations in scattering properties could explain its poorer performance.

Neither transmittance retrieval method presented on Figure 3 accounted for measured ice thickness. This choice was made to test the algorithm in the case where ice thickness data are unavailable. We also tested the case where such ice thickness data would be available (Appendix B). There we did a similar analysis, but with an exponential attenuation factor incorporating the effect of sea ice thickness, for both methods. This correction has a small positive effect on the overall trend. Indeed, we note a slightly higher R-squared ($R^2$=0.795), slightly lower root mean square error (RMSE=0.012) and slightly lower mean bias magnitude (MB=-0.0024) with the spectral-albedo inversion. However, it significantly improves estimation for transmittance above $10^{-2}$ or, equivalently, for $\tau'_{1,2}$ below 20. This threshold $\tau'_{1,2}$ is consistent with findings based on simulations.

### 3.3 Effect of the Scattering Properties on Transmittance

To further investigate why the spectral albedo-based retrievals perform better than BL estimates, we present an analysis of the links between transmittance, reduced optical thickness $\tau'_{1,2}$ of layers 1 and 2 and snow depth. Figure 4 illustrates transmittance estimated from (a) $\alpha_\lambda$-inversion and (b) the BL method as a function of each method's main driver of transmittance. For $\alpha_\lambda$-inversion, this driver is $\tau'_{1,2}$, calculated from vertically resolved scattering properties of layers 1 and 2 (see Eq.7). To obtain an R-squared value, $\tau'_{1,2}$ is represented on a logarithmic scale. For the BL method, this driver is measured geometric snow depth. These estimations are from the same five campaigns as presented in Figure 3. To illustrate the effect of the microstructure-driven scattering properties on the accuracy of transmittance, the reduced scattering coefficient $b'_{1,inv}$ of layer 1, obtained from $\alpha_\lambda$-inversion, is depicted by a colorscale.

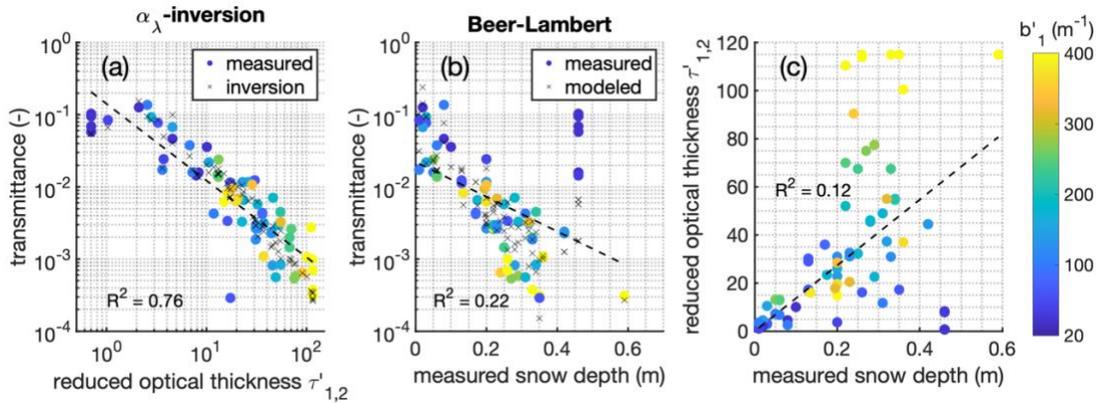

Figure 4: Transmittance as a function of the main driver for each respective method: The driver for (a) the square root of the reduced optical depth of layer 1 and 2 inverted from spectral albedo and for (b) is measured snow depth. (c) The correlation between both drivers is evaluated. The upper layer reduced scattering coefficient is depicted by the colorbar.

Comparing Figure 4 a-b, one can notice that measured transmittance much better relates to $\tau'_{1,2}$ than to snow depth . This is supported by $\tau'_{1,2}$ having a far better R-squared value in regard to transmittance. This can also be observed from the fact that the relation to snow depth display a greater amount of points that are significantly off the trend. Interestingly, in the relation to snow depth, the transmittances that are significantly higher than the trend have very low $b'_{1,inv}$ and the transmittances that are significantly lower than the trend have very high $b'_{1,inv}$. No such phenomenon is observed in the relation to reduced optical depth. Thus, we argue that this is because reduced optical depth directly accounts for the effect of scattering while snow depth does not.

As observed on Figure 4 c, the relationship between snow depth and $\tau'_{1,2}$ is not simple. This relation does not follow a linear trend arguably because $\tau'_{1,2}$ accounts for the variability of snow and sea ice microstructure-driven scattering properties, whereas snow depth does not. Indeed, all the points away from the trend have very high or very low $b'_{1,inv}$.

By construction, because they well capture microstructure and associated scattering properties variation in addition to geometric layer thicknesses, spectral albedo-based retrievals capture transmittance variations much better than BL-based retrievals, which mainly counts on snow geometric depth.

## 4 Conclusion

We presented a method to estimate transmittance under first-year snow-covered sea ice relying solely on a spectral albedo measurement. The inversion algorithm relies on its ability to invert the vertically resolved scattering properties of snow and sea ice above the freeboard to estimate transmittance.

We have shown, using field measurements from five campaigns, that spectral albedo transmittance retrievals seem to better account for the actual vertical optical structure of the upper snow-ice system than simpler BL attenuation models based mainly on snow geometric depth.

This gain in performance in comparison to the BL method could potentially be even more significant if the method was upscaled either using an airborne platform or a satellite. This is because errors on remote surface albedo estimates are generally much smaller than those in remote snow depth estimates.

Many remotely applicable improvements could make the spectral albedo inversion technique even more reliable. As examples: adding information on sea ice thickness and sun zenithal angle, optimizing the spectral range of the inversion based on sea ice type, relying on more sophisticated spectra comparison methods or implementing a quality criterion on the inversion. With such room for improvements, the spectral albedo inversion method has the potential to provide pan-Arctic PAR maps useful for the assessment of under-ice primary production; and with some creativity, provide other key sea ice related estimates.

**Appendix A**

Vertically resolved scattering properties obtained from spectral albedo inversion on first-year snow-covered sea ice during the QikIce2023 campaigns were validated to in situ scattering properties. Figure A1 compares the vertical profile of the reduced scattering coefficient for both methods at every site.

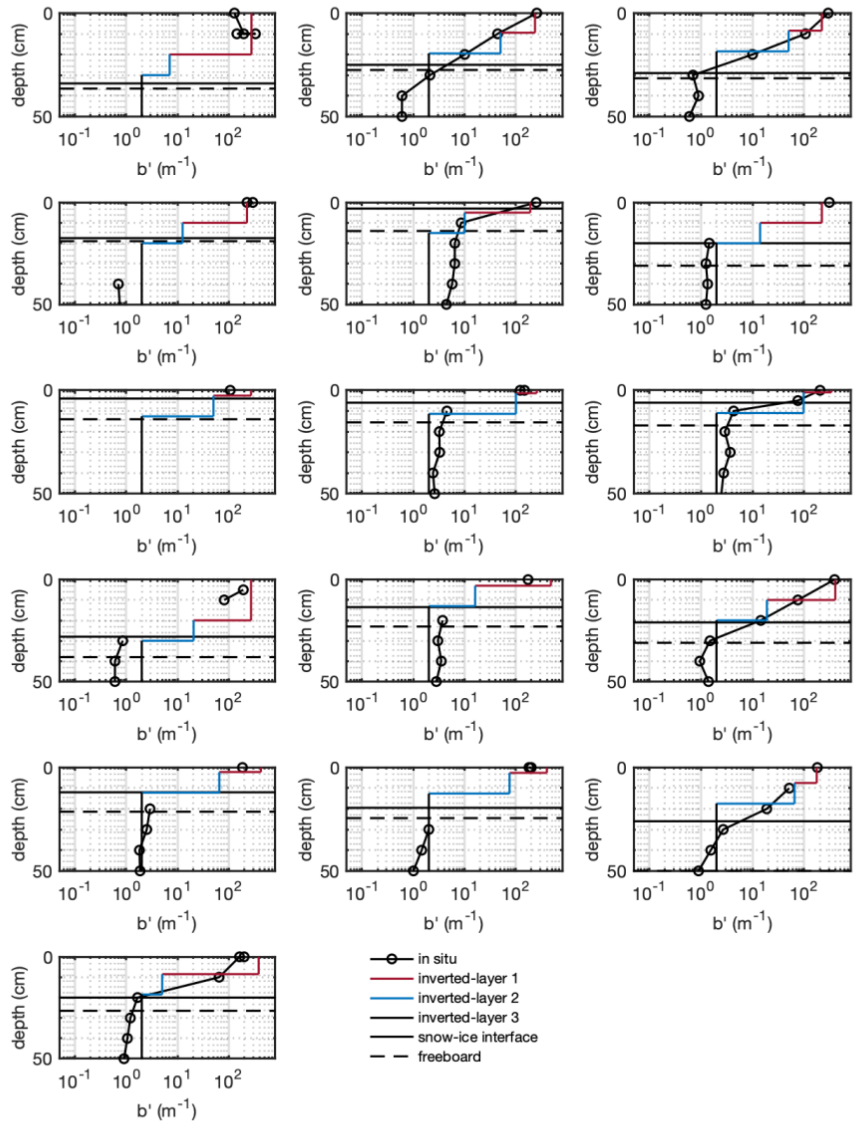

Figure A1: Vertically resolved scattering properties validation from QikIce2023: comparison between three-layer reduced scattering coefficient profiles b' obtained from albedo inversion and in situ b' obtained with an active reflectance probe. The depth level z=0 cm corresponds to the air-snow interface.

**Appendix B**

We compared transmittance estimated from albedo inversion to transmittance estimated from Beer-Lambert method. Figure B1 presents the same results as Figure 3 presented in the main text with the exception that, instead of fixing interior sea ice attenuation, it is corrected for both methods to meet actual sea ice attenuation based on sea ice thickness and a Beer-Lambert correction factor. Accounting for actual sea ice attenuation slightly improves R-square from 0.771 to 0.794 for the albedo inversion method. This improvement is mostly observed at high transmittance (T~$10^{-1}$) and low optical thickness values.

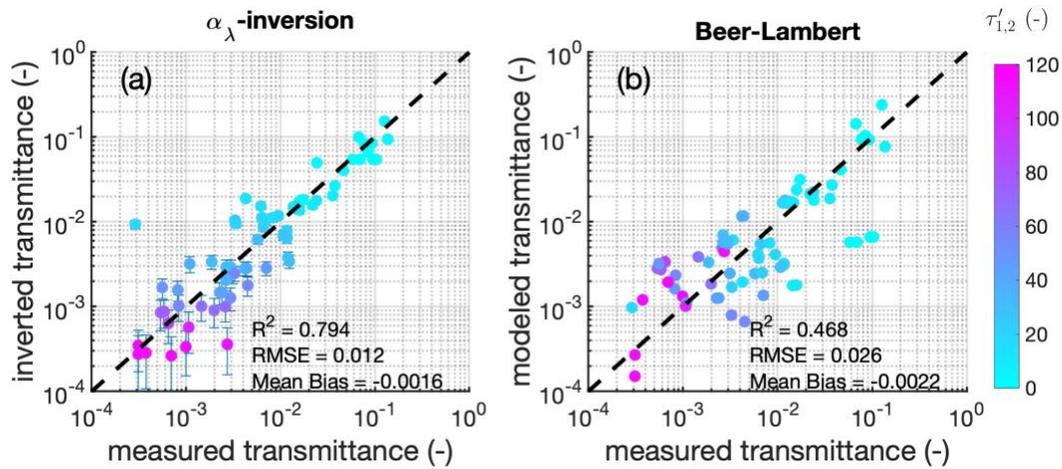

Figure B1: Field validation of transmittance under snow-covered sea ice estimated from (a) inversion of spectral albedo and (b) two-level Beer-Lambert method based on broadband albedo and snow depth. In this annexe, attenuation by sea ice is corrected using a Beer-Lambert's factor of $e^{-k_i \cdot \Delta h_i}$ to meet actual sea ice attenuation. The colorbar is reduced optical depth obtained from inversion of spectral albedo.

## Acknowledgments

We would like to thank Guislain Bécu, Raphael Larouche, Béatrice Lessard-Hamel, Manon Gibaud, Joannie Ferland, Gabriel Lapointe and Marie-Hélène Forget for their precious logistic support during the field deployment in the Arctic and throughout the whole process. Félix Levesque-Desrosiers for the insightful discussions.

Finally, the research project was supported by the Canada Excellence Research Chair on Remote Sensing of Canada's new Arctic frontier, Discovery grant no. RGPIN-2020-06384 to Marcel Babin, the SMAART program funded by the Collaborative Research and Training Experience Program of the Natural Sciences and Engineering Research Council of Canada, the Québec Océan program and the Sentinel North program of Université Laval, made possible, in part, thanks to funding from the Canada First Research Excellence Fund.

## Open Research

To be provided